\documentclass[12pt]{article}
\usepackage{amsmath}
\usepackage{graphicx}
\usepackage{amssymb}
\usepackage{a4wide}
\usepackage{bbm}
\usepackage{latexsym}
\date{}

\newcommand{\Eqref}[1]{Eq.~\eqref{#1}}

\title{\sf Scalar Casimir-Polder forces for uniaxial corrugations}
\author{Babette D\"{o}brich${}^{1,2,3}$, Maarten DeKieviet${}^3$, and Holger
  Gies${}^{1,2}\footnote{E-mail: babette.doebrich@uni-jena.de, maarten.dekieviet@physik.uni-heidelberg.de,
gies@tpi.uni-jena.de}$\\
  \\
  {\small \it ${}^1$ Theoretisch-Physikalisches Institut,
    Friedrich-Schiller-Universit\"at Jena}\\
  {\small \it Max-Wien-Platz 1, D-07743 Jena, Germany}\\
  {\small \it ${}^2$ Institut f\"ur Theoretische Physik,
    Universit\"at Heidelberg}\\
  {\small \it Philosophenweg 16, D-69120 Heidelberg, Germany}\\
  {\small \it ${}^3$ Physikalisches Institut,
    Universit\"at Heidelberg}\\
  {\small \it Philosophenweg 12, D-69120 Heidelberg, Germany}\\
}

\begin{document}

\maketitle

\begin{abstract}
  We investigate the Dirichlet-scalar equivalent of Casimir-Polder forces
  between an atom and a surface with arbitrary uniaxial corrugations. The
  complexity of the problem can be reduced to a one-dimensional Green's
  function equation along the corrugation which can be solved numerically. Our
  technique is fully nonperturbative in the height profile of the
  corrugation. We present explicit results for experimentally relevant
  sinusoidal and sawtooth corrugations. Parameterizing the deviations from the
  planar limit in terms of an anomalous dimension which measures the power-law
  deviation from the planar case, we observe up to order-one anomalous
  dimensions at small and intermediate scales and a universal regime at larger
  distances. This large-distance universality can be understood from the fact
  that the relevant fluctuations average over corrugation structures smaller
  than the atom-wall distance.

\end{abstract}

\section{Introduction}

Casimir forces \cite{Casimir:dh} between mesoscopic or macroscopic objects as
well as Casimir-Polder forces \cite{CasimirPolder} between an atom and a
surface can be attributed to a reordering of fluctuations in the quantum
vacuum. Recent years have witnessed considerable progress in measuring these
forces \cite{Lamoreaux:1996wh,Mohideen:1998iz,DeKieviet3,Decca:2007yb}, paving
the way for future application in micro- and nanomechanical engineering and
single-atom manipulation.

Particularly for such applications, standard calculational techniques for
simple flat surfaces are insufficient and a profound understanding of the
influence of geometry on these quantum forces is required. In fact, since
fluctuations occur on all momentum or length scales, quantum forces are
strongly affected by the global properties of a given system. From a technical
perspective, global properties such as geometry or curvature dependencies
generally require a full understanding of the fluctuation spectrum in a given
configuration and cannot be dealt with by perturbative expansions with respect
to a small geometry parameter.

Therefore, a variety of new field-theoretical methods for understanding and
computing fluctuation phenomena have been developed in the past few years,
such that early phenomenological recipes, such as the proximity force
approximation (PFA) \cite{pft1} have been overcome by now. Apart from exact
results in certain asymptotic limits \cite{Feinberg,Balian:1977qr} and more
controllable approximation techniques \cite{semicl,Scardicchio:2004fy},
field-theoretical worldline methods have lead to efficient algorithms for
Casimir energies \cite{Gies:2003cv,Gies:2005ym,Gies:2006bt}. In addition, approaches based on
scattering theory have proved most successful for finding new exact solutions
and efficient computation schemes
\cite{Bulgac:2005ku,Kenneth:2006vr,Emig:2006uh,Bordag:2006vc,Rodrigues:2006ku,Emig:2007cf,Milton:2007gy,Milton:2008vr},
see \cite{Milton:2008st} for a recent review.
New results with direct mode summation have been obtained in
\cite{Mazzitelli:2006ne}; numerical tools based on brute-force discretization
have been used in \cite{Joannopoulos}.

Many of these approaches can be directly linked with a
constrained-functional-integral formulation, as first introduced in
\cite{Bordag:1983zk} for the parallel-plate case and further developed for
corrugated surfaces in \cite{Emig:2001dx} and lastly extended to general dispersive
forces between structured media \cite{Emig:2003eq}. For Casimir forces
involving corrugated surfaces, results based on a perturbative expansion in
the height profile have been obtained recently \cite{CaveroPelaez:2008tj}. 

Whereas most of these approaches have been mainly applied to Casimir forces
between extended objects, the Casimir-Polder force between an atom and a
surface of general shape has not been so widely studied. An estimate of the
influence of surface roughness has been given in \cite{bezerra}, where an
additivity approximation has been perturbatively expanded in the roughness
amplitude.  In \cite{dalvit}, it was pointed out that large geometry
corrections to the Casimir-Polder force should be observable for atoms in
front of a corrugated surface, potentially visible in experiments with cold
atoms near a surface. These studies were based on a perturbative
analysis in the height profile. If the height-profile parameters are of the same
order as the atom-wall distance, however, geometry plays a much more dominant 
role and perturbative theory is expected to be no longer
appropriate. Whereas cold atom gas experiments become increasingly difficult in
this regime, Casimir-Polder force measurements can be well controlled in this
regime. In particular the atomic beam spin echo technique introduced in 
\cite{DeKieviet1, DeKieviet2, DeKieviet3} has recently demonstrated high resolution 
access to some of the fundamental issues concerning Casimir interactions. 
Experimental results on geometrical effects will be presented elsewhere
shortly \cite{Warring}.  

This work is devoted to a nonperturbative study of Casimir-Polder forces near
a surface with uniaxial corrugation. Our method is based on the
constrained-functional-integral approach which facilitates to map the core
part of the problem onto a one-dimensional Green's function problem along the
direction of nontrivial curvature of the surface. As this Green's function
problem involves singular kernels, we identify an appropriate representation
which is accessible to stable and efficient numerical tools.  For simplicity,
we consider a fluctuating scalar field obeying Dirichlet boundary conditions
instead of the full electromagnetic field. Hence, our results should not be
directly applied to realistic atom-wall studies; they are applicable to
Casimir configurations in ultracold-gas systems as suggested in
\cite{Roberts:2005}. Our method is not restricted to Dirichlet
scalars and can straightforwardly be generalized to the electromagnetic case.

The paper is organized as follows: In Sect.~\ref{sec:scalar_ft_boundaries}, we
review the treatment of Casimir forces within scalar QFT and give the general
Casimir energy for two Dirichlet surfaces. Within this formulation, we then
evaluate the Casimir energy in Sect.~\ref{sec:planar} for a sphere and a
planar surface in the limit $\frac{r}{H} \ll 1$ to first order
explicitly. This limit defines the scalar analogue of the Casimir-Polder
force. In Sect.~\ref{sec:corrugated}, we extend our technique to the case of
uniaxially corrugated surfaces. In Sect.~\ref{sec:sinus}, we present numerical
results for the scalar Casimir-Polder potential for a
sinusoidally shaped surface as well as for a saw tooth profile, both of which
have been used in experiments \cite{Warring}. Conclusions are given in Sect.~\ref{sec:conclusions}.

\section{Scalar field theory with boundaries\label{sec:scalar_ft_boundaries}}

The Casimir energy of a system is given by the shift of the ground state
energy caused by the presence of boundaries imposing constraints on the
fluctuating field. For calculating Casimir forces between disconnected
bodies, only the Casimir interaction energy is relevant. The latter
corresponds to that part of the ground state shift which depends on the
relative position and orientation of all bodies. The Casimir self-energies of single bodies are
irrelevant for the Casimir force. 

Here, we investigate the Casimir interaction energy induced by fluctuations of
a scalar field $\phi$. We follow the constrained-functional-integral approach
\cite{Bordag:1983zk}, which is introduced in the following in a
brief and simplified manner, see \cite{Bordag:1983zk,Emig:2001dx,PhysRevA.67.022114,Emig:2007cf}
for a more detailed discussion and generalizations. We start
from the associated Euclidean generating functional,
\begin{equation}
  \mathcal{Z}=\int\mathcal{D}\phi\exp\left(-\frac{1}{2\hbar}\int
  d^{4}x\left(\partial\phi(x)\right)^{2}\right).
\end{equation} 
The Casimir interaction energy of a system bounded by two surfaces, whose
relative position is specified (e.g., by a (mean) separation $H$),
is then given by
\begin{equation}
  E(H)=-\frac{\hbar c}{T_{\text
      E}}\ln\frac{\mathcal{Z}_{\text{B.C.}}}{\mathcal{Z}_{\infty}}\
  ,\label{general_cas_energy}
\end{equation} 
where $\mathcal{Z}_{\text{B.C.}}$ stands for the generating functional of
the fluctuating field obeying the system's boundary conditions, whereas
$\mathcal{Z}_{\infty}$ represents the case of infinite separation
between the objects, i.e., $H\rightarrow\infty$. In this way, irrelevant
Casimir self-energies are subtracted.  The length in Euclidean time direction
is denoted by $T_{\text{E}}$.

The boundary conditions for the fields are implemented by insertion of a
$\delta$ functional constraint into $\mathcal{Z}$.  For the case of Dirichlet
boundary conditions, the corresponding $\delta$ functional can be represented
by product of $\delta$ functions
$\delta\left(\phi\left(x_{\alpha}\right)\right)$ for all 4-vectors
$x_{\alpha}$ pointing onto a surface $S_{\alpha}$; here, $\alpha$ labels
multiple disjoint surfaces.  Hence, $\mathcal{Z}_{\text{B.C.}}$ for Dirichlet
boundaries in the case of two plates is given by
\begin{equation}
\mathcal{Z}_{D}=\int\mathcal{D}\phi
\prod_{\alpha=1}^{2}\prod_{x_{\alpha}}\delta\left(\phi(x_{\alpha})\right)
\exp\left(-\frac{1}{2\hbar}\int
  d^{4}x\left(\partial\phi(x)\right)^{2}\right)
.\label{Z_dirichlet}
\end{equation} 
In order to evaluate the integral over the fields $\phi$, a Fourier
representation is used for the $\delta$ functional with the help of auxiliary
fields that have support only on the surfaces $S_{\alpha}$.\footnote{Drawing
  the analogy to the electromagnetic case, the auxiliary fields can be thought
  of as charged sources which enforce the boundary conditions by means of
  their coupling to the fluctuating field \cite{Emig:2007cf}.} 
First performing the Gau\ss ian integral over $\phi$ leaves us -- apart
from an irrelevant factor -- with another Gau\ss ian integral for the
auxiliary fields which can also be carried out, yielding for the Casimir
interaction energy,
\begin{equation}
E(H)  =  \frac{\hbar c}{T_{\text{E}}}\frac{1}{2}\mathsf{Tr}
\ln(\mathcal{M}_{\infty}^{-1}\mathcal{M})\label{E_ln_M_overM} ,
\end{equation}
where $\mathcal{M}$ denotes a matrix whose entries are the propagators of the fluctuating scalar, i.e.,
the Green's function of the Laplacian in this case,
\begin{equation}
\mathcal{M}_{\alpha\beta}(x,x')=\frac{1}{4\pi^{2}}\frac{1}{(x_\alpha-x_\beta')^{2}}
,\label{greens_fct_for_del_square}
\end{equation}
with four-vectors $x$ and $x'$ pointing onto the surfaces $S_{\alpha}$ and
$S_{\beta}$, respectively. At the same time, $\mathcal{M}$ is the inverse of
the auxiliary-field propagator on the surfaces. Complications of
Eq.~\eqref{E_ln_M_overM} are hidden in the fact that the trace has to be
evaluated over the spacetime hypersurface of the boundaries only, as specified
by the support of the auxiliary fields. As the surfaces can be curved, this
requires appropriate metric factors for the spatial integration measures 
(see below).  The normalizing operator $\mathcal{M}_{\infty}$ corresponds to the
propagator at infinite separation of the surfaces and takes
care of the self-energy subtraction.  For the case of two surfaces,
$\mathcal{M}_{\infty}^{-1}\mathcal{M}$ can be written as
\begin{equation}
\mathcal{M}_{\infty}^{-1}\mathcal{M}=\left(\begin{array}{cc}
1 & \mathcal{M}_{11}^{-1}\mathcal{M}_{12}\\
\mathcal{M}_{22}^{-1}\mathcal{M}_{21} & 1\end{array}\right)
\label{M_M_inf} ,
\end{equation}
where the matrix components distinguish between spacetime arguments "living" on the two
different surfaces, i.e., $\alpha,\beta=1,2$. In the matrix product, an
integration over the connecting intermediate spacetime points is implicitly
understood. Due to the normalization with $\mathcal{M}_{\infty}$, only the
off-diagonal elements are non-trivial, as they contain information about the
propagation of fluctuations between the two different surfaces.

In order to compute the Casimir energy from
Eq.(\ref{E_ln_M_overM}) explicitly, we introduce the off-diagonal matrix
$\Delta \mathcal M$, 
\begin{equation}
\mathcal{M}_{\infty}^{-1}\mathcal{M}= \boldsymbol{1}+ \left(\begin{array}{cc}
0 & \mathcal{M}_{11}^{-1}\mathcal{M}_{12}\\
\mathcal{M}_{22}^{-1}\mathcal{M}_{21} & 0\end{array}\right) \equiv
\boldsymbol{1}+\Delta\mathcal{M} ,
\end{equation}
such that a series expansion of the logarithm in Eq.(\ref{E_ln_M_overM})
yields
\begin{equation}
E(H) = -\frac{\hbar
  c}{T_{\mathrm{E}}}\frac{1}{2}\sum_{n=1}^{\infty}\frac{1}{2n}
\mathsf{Tr}\left(\Delta\mathcal{M}^{2n}\right)\label{E_tr_sum}
.\end{equation}
Due to its off-diagonal structure, only even powers of
$\Delta\mathcal{M}$ contribute to the Casimir energy.  Performing the discrete
trace over the matrix entries then yields by the cyclicity of the trace
\begin{equation}
E(H) = -\frac{\hbar
  c}{T_{\mathrm{E}}}\frac{1}{2}\sum_{n=1}^{\infty}\frac{1}{n}
\mathsf{Tr}\left(\left(\mathcal{M}_{11}^{-1}\mathcal{M}_{12}
\mathcal{M}_{22}^{-1}\mathcal{M}_{21}\right)^{n}\right)\label{E_tr_cycle} .
\end{equation}
Thus, the Casimir energy can be understood as a sum over all generic "chains"
of correlators between the surfaces. Starting from a generic point on surface
$S_{1}$, it is summed over all possible auxiliary-field or source correlators
on the surface itself, followed by all possible $\phi$ propagations to the
second surface $S_{2}$, then generic source propagations on $S_{2}$ itself
and all thinkable ways back to $S_{1}$ again.  As a last step, these Casimir contributions are summed over all
possible starting points. Within the language of scattering theory,
$\mathcal{M}_{\alpha\alpha}^{-1}$ is related to the $T$ matrix associated with
scattering off the surface $\alpha$ \cite{Emig:2007cf}.

Equation \eqref{E_tr_cycle} is a useful starting point for explicit
computations for a given configuration. For instance for two parallel
Dirichlet planes, the trace turns out to be proportional to $1/(nH)^3$, yielding
the scalar analogue to Casimir's celebrated result upon summation over
$n$. The present formalism can straightforwardly be generalized to Neumann or
electromagnetic boundary conditions by insertion of the corresponding $\delta$
functional constraint. The formalism has also been worked out for dielectric
materials and fluctuations in media from which Lifshitz theory \cite{lifshitz}
can immediately be derived \cite{Emig:2003eq}. Arbitrarily curved surfaces
require the knowledge of $\mathcal{M}_{\alpha\alpha}^{-1}$ for both surfaces,
which is equivalent to working out the $T$ matrix (or $S$ matrix) of the
corresponding scattering problem. Moreover, an efficient means to carry out
the remaining summation and integrations is needed as well.

\section{Scalar Casimir-Polder potential for a planar
  surface\label{sec:planar}}

The Casimir-Polder force between an atom and a plane wall at a
distance $H$ is easily derived from Lifshitz theory for two parallel
plates. Consider one of the plates as consisting of a dilute dielectric with a
dielectric permittivity $\epsilon= 1+4\pi \alpha N+\mathcal O(N^2)$. Here,
$\alpha$ denotes the polarizability of the atoms in the surface and $N$ is the
number of atoms per unit area in the surface. Expanding the Casimir force from
Lifshitz theory to first order in $N$ yields the Casimir-Polder force between
one of the atoms in the dielectric and the opposite wall as the prefactor at
order $N$, see, e.g., \cite{babb}. This approach, however, cannot
simply be applied to the case of an atom near a corrugated surface,
since the prefactor at order $N$ is an average over all possible atom positions
at a given mean distance $H$ above the corrugated surface. Hence, the important
information about the dependence of the force on lateral coordinates is not
available in this manner.

This information can be extracted following a different strategy. As the
Casimir-Polder force arises from the atom's polarizability, i.e., the dipole
transition induced by the fluctuating field, we can represent the atom by a
compact surface with suitable polarization properties. The simplest case is, of
course, a small sphere of radius $r$. In the limit $r\ll H$, the lowest
non-vanishing inducible multipole moment, i.e., the dipole polarizability in
this case, dominates the Casimir interaction between the small sphere and the corrugated
surface. Hence, identifying this dipole polarizability of the sphere with that
of an atom yields the Casimir-Polder law in this limit. 

In this section, we proceed in exactly this fashion for the Dirichlet scalar
case, i.e., we define the Casimir-Polder law in this case as the Casimir force
between a small sphere of radius $r$ and a corrugated wall at (mean) distance $H$ in the
limit of $r\ll H$. Here the lowest multipole contribution of a small sphere is
that of a monopole excitation, hence the resulting power law \cite{Gies:2005ym,Bulgac:2005ku} with distance $H$
will be different to that of the electromagnetic case
, but it is straight forward to generalize this approach and
computational strategy to the electromagnetic case. 

As an exercise, we start with the planar case, identifying the planar surface
with $S_1$ parametrized by the spatial coordinate $x_3 =0$. The center of the
sphere (surface $S_2$) of radius $r$ is located at $x_{1}=x_{2}=0$,
$x_{3}=H$. The singularity structure of the involved propagators is not
entirely trivial and special care has to be taken to choose the proper order of limits and integrations. Of the four correlators occurring in Eq.(\ref{E_tr_cycle}), we
begin with the contraction of the inverse propagator on the plane with the
propagator between the surfaces.  

Specifying Eq.(\ref{greens_fct_for_del_square}) for the plane $S_{1}$ results
in $\mathcal{M}_{11}= 1/ [4\pi^{2}(\underline{x}-\underline{x}')^{2}]$, where
$\underline{x}=(x_{0},x_{1},x_{2})=(x_{0},\vec{x}_{\parallel})$, i.e., the
$x_3$ coordinate perpendicular to the plane drops out. By Fourier
transforming the propagator to postion space, we obtain the functional
inverse,
\begin{equation}
\mathcal{M}_{11}^{-1}=2\sqrt{-\underline\nabla^{2}}\label{m11inverse_flach} ,
\end{equation}
where $\underline\nabla=(\partial_{0},\partial_{1},\partial_{2})$. This
operator can immediately be contracted with the correlator between the plane
and the sphere yielding
\begin{equation}
\int_{\underline{x}''}\mathcal{M}_{11}^{-1}(\underline{x};\underline{x}'')\mathcal{M}_{12}(\underline{x}'';x')
=\frac{1}{\pi^{2}}\frac{x_{3}'}{(x_{3}'^{2}
+(\underline{x}-\underline{x}')^{2})^{2}}\equiv\Delta\mathcal{M}_{12}(\underline{x};x')
.\label{deltam_12_par_plates}
\end{equation} 
Since the setup is translationally invariant in time direction, it is
expedient to Fourier transform the propagators to frequency space with respect
to the time coordinate. The singularity structure can be well controlled using
a propertime integral representation (with propertime $S$). In addition, we
parametrize the surface of the sphere $S_2$ by spherical coordinates.  The
combined propagator then reads
\begin{multline}
\Delta\mathcal{M}_{12}(\zeta;\vec{x}_{\parallel};\Omega)
=\frac{1}{\pi^{\frac{3}{2}}}(H+r\cos\theta)\int_{0}^{\infty}dS
\sqrt{S}\exp\left(-\frac{\zeta^{2}}{4S}\right)\\
\times \exp\left(-\left[(x_{1}-r\cos\phi\sin\theta)^{2}
+(x_{2}-r\sin\phi\sin\theta)^{2}+(H+r\cos\theta)^{2}\right]S\right),\\
 \label{deltaM12_prop_time}
\end{multline}
where $\zeta$ denotes the imaginary frequency, the lateral coordinates on the
plane are given by $\vec{x}_{\parallel}$, and $\Omega$ summarizes the
azimuthal ($\theta$) as well as the polar ($\phi$) angle on the sphere.

Next, the inverse propagator $\mathcal{M}_{22}^{-1}$ on the sphere $S_{2}$ can
be determined from $\mathcal{M}_{22}^{-1} \mathcal{M}_{22} = \mathbbm{1}$
which in terms of spherical coordinates reads
\begin{equation}
r^{2}\int_{\Omega'}\mathcal{M}_{22}^{-1}(\zeta;\Omega;\Omega')
\mathcal{M}_{22}(\zeta;\Omega';\Omega'')
=\frac{1}{r^{2}}\delta(\phi-\phi'')\delta(\cos\theta-\cos\theta'') ,
\label{m22_inv_bestimmgl_polar}
\end{equation}
with $\mathcal{M}_{22}$ given by \Eqref{greens_fct_for_del_square}, and
$\int_{\Omega}=\int_{0}^{2\pi}\mathrm{d}\phi\int_{-1}^{1}\mathrm{d}(\cos\theta)$.
This equation can be solved for $\mathcal{M}_{22}^{-1}$ by expanding all
quantities in spherical harmonics $Y_{lm}(\Omega)$. As mentioned above, only
the monopole order $l=0=m$ is relevant here, yielding 
\begin{equation}
\mathcal{M}_{22}^{-1}(\zeta)\Big|_{\text{monopole}}
=\frac{1}{4\pi}\frac{\left|\zeta\right|\exp(r|\zeta|)}{r^{2}\sinh(r|\zeta|)}
.\label{M22inv_kugel}
\end{equation} 
The last propagator in the chain (Eq.(\ref{E_tr_cycle})) evaluated on the
surfaces $S_{1}$ and $S_{2}$ reads, again in propertime representation,
\begin{equation}
\mathcal{M}_{21}(\zeta;\Omega;\vec{x}_{\parallel})=\int_{0}^{\infty}\frac{dT}{\sqrt{4\pi
    T}}\, e^{-\frac{\zeta^{2}}{16\pi^{2}T}}
e^{-4\pi^{2}\left[(r\cos\phi\sin\theta-x_{1})^{2}+(r\sin\phi\sin\theta-x_{2})^{2}
+(H+r\cos\theta)^{2}\right]T} .\label{m21_prop_time}
\end{equation}
With this, all correlators needed for the calculation of the Casimir energy
between the sphere and the plane to monopole order are at hand. For the
Casimir-Polder limit, it suffices to consider the lowest-order term of the $n$
sum in \Eqref{E_tr_cycle}, as discussed below. This term reads
\begin{equation}
\mathsf{tr}\left(\Delta\mathcal{M}_{12}\mathcal{M}_{22}^{-1}\mathcal{M}_{21}\right)=\\
\frac{T_{\text{E}}}{2\pi}\int_{\zeta}\int_{\vec{x}_{\parallel}}\int_{\Omega}
\int_{\Omega'}r^{4}\Delta\mathcal{M}_{12}(\zeta;\vec{x}_{\parallel};\Omega)
\mathcal{M}_{22}^{-1}(\zeta)\mathcal{M}_{21}(\zeta;\Omega';\vec{x}_{\parallel})
.\label{trMMM_z_xp_o_o}
\end{equation}
Let us now rescale all dimensionful quantities by the scale set by the
sphere-plate distance $H$, i.e.,
$\vec{x}_{\parallel}\rightarrow\tilde{\vec{x}}_{\parallel}H$,
$\zeta\rightarrow\frac{\tilde{\zeta}}{H}$, and consequently also the
propertime parameters $S\rightarrow\frac{\tilde{S}}{H^{2}}$ and
$T\rightarrow\frac{\tilde{T}}{H^{2}}$.  In the limit $\frac{r}{H}\ll1$, the
inverse propagator on the sphere \eqref{M22inv_kugel} can be
expanded as
\begin{equation}
\mathcal{M}_{22}^{-1}(\tilde{\zeta})=\frac{1}{4\pi}\frac{1}{r^{3}}\left(1
  +\mathcal{O}\left(\frac{r}{H}|\tilde{\zeta}|\right)\right).\label{m_expanded}
\end{equation} 
Even though $\zeta$ ranges from 0 to $\infty$, the $\tilde\zeta$ integral
receives its dominant contributions from $\tilde\zeta=\mathcal {O}(1)$, such
that the order estimate of \Eqref{m_expanded} is meaningful. The inverse
propagator $\mathcal{M}_{22}^{-1}$ thus becomes independent of the imaginary
frequency to lowest order. 

Collecting all the dimensionful factors from the rescaling, the trace
Eq.(\ref{trMMM_z_xp_o_o}) turns out to be of order
$\mathcal{O}\left(\frac{r}{H^{2}}\right)$ and the remaining dimensionless
integrals are a pure function of $r/H$. In the limit $r\ll H$,
$\Delta\mathcal{M}_{12}(\tilde{\zeta};\tilde{\vec{x}}_{\parallel};\Omega)$ and
$\mathcal{M}_{21}(\tilde{\zeta};\Omega';\tilde{\vec{x}}_{\parallel})$ become
independent of $\Omega$ and $\Omega'$, such that the two
solid-angle integrations just contribute a factor of $16\pi^{2}$.
We then obtain for Eq. (\ref{trMMM_z_xp_o_o}):
\begin{equation}
\mathsf{tr}\left(\Delta\mathcal{M}_{12}\mathcal{M}_{22}^{-1}\mathcal{M}_{21}\right)
=T_{\text{E}}\frac{r}{H^{2}}2\int_{\tilde{\zeta}}\int_{\tilde{\vec{x}}_{\parallel}}
\Delta\tilde{\mathcal{M}}_{12}(\tilde{\zeta};\tilde{\vec{x}}_{\parallel})
\tilde{\mathcal{M}}_{21}(\tilde{\zeta};\tilde{\vec{x}}_{\parallel})
+\mathcal{O}\left(\frac{r^{2}}{H^{3}}\right)
,\label{tr_plate_kugel_dimensionless}
\end{equation} 
with the associated dimensionless propagators
\begin{eqnarray}
\Delta\tilde{\mathcal{M}}_{12}(\tilde{\zeta};\tilde{\vec{x}}_{\parallel}) 
& = &
\frac{1}{\pi^{\frac{3}{2}}}\int_{0}^{\infty}\mathrm{d}\tilde{S}\sqrt{\tilde{S}}
\,e^{-(1+\tilde{\vec{x}}_{\parallel}^{2})\tilde{S}}
\exp\left(-\frac{\tilde{\zeta}^{2}}{4\tilde{S}}\right)\nonumber \\
\tilde{\mathcal{M}}_{21}(\tilde{\zeta};\tilde{\vec{x}}_{\parallel}) 
& = &
\frac{1}{2\sqrt{\pi}}\int_{0}^{\infty}\frac{\mathrm{d}\tilde{T}}{\tilde{T}}
\,e^{-4\pi^{2}(1+\tilde{\vec{x}}_{\parallel}^{2})\tilde{T}}
\exp\left(\frac{\tilde{\zeta}^{2}}{16\pi^{2}\tilde{T}}\right)
.\label{deltaM12_M21_resc_proptime}
\end{eqnarray}
Finally, the remaining integrations in the trace expression can be performed
straightforwardly in the following order: First, we perform the integral over
$\tilde{\zeta}$ which is purely Gau\ss ian, then the integral over
$\tilde{\vec{x}}_{\parallel}$ is done using polar coordinates. At last, the
integration over the variables of the proper time representation $\tilde{T}$
and $\tilde{S}$ is performed.  The result for the lowest-order trace term then 
reads
\begin{equation}
  \mathsf{tr}\left(\Delta\mathcal{M}_{12}\mathcal{M}_{22}^{-1}\mathcal{M}_{21}\right)
=T_{\text{E}}\frac{1}{4\pi}\frac{r}{H^{2}}+\mathcal{O}\left(\frac{r^{2}}{H^{3}}\right)
\label{trace_scalar_cp_flach}
  .
\end{equation} 
As higher orders in the $n$ sum involve more propagators between the sphere
and the plate and consequently further powers of $H$ in the denominator, only
the $n=1$ term survives in the Casimir-Polder limit. Using
Eq. (\ref{E_tr_cycle}), the energy between the sphere and the plane
consequently yields in the Casimir-Polder limit $r\ll H$
\begin{equation}
  E(H)=-\frac{\hbar
    c}{8\pi}\frac{r}{H^{2}}+\mathcal{O}\left(\frac{r^{2}}{H^{3}}\right)
  ,\label{potential_scalar_cp_flach}
\end{equation}  
which agrees with
\cite{Gies:2005ym,Gies:2006bt,Bulgac:2005ku,emig_scalar_dirichlet,MaiLamRey}.

As we have seen, the spatial integrations over the surface of the sphere
$S_{2}$ become trivial in the Casimir-Polder limit. Due to this fact, the
integrations over the remaining lateral coordinates $\vec{x}_{\parallel}$ on
the plate (cf. Eq. \eqref{tr_plate_kugel_dimensionless}) could also have been
performed in momentum space since the flat plate itself is translationally
invariant along these directions. However, in the following section we will
extend our investigations to surfaces which are uniaxially structured along
the direction $x_{1}$. For this purpose, it is expedient to Fourier
transform Eqs.(\ref{tr_plate_kugel_dimensionless}) and
(\ref{deltaM12_M21_resc_proptime}) only with respect to the 2-component
$p_{2}$ to momentum space. Due to a remaining Lorentz invariance in time and
$x_2$ direction, the integrand only depends on the combination of momenta
$\tilde{q}=\sqrt{\tilde{p}_{2}^{2}+\tilde{\zeta}^{2}}$, such that we obtain
\begin{equation}
\mathsf{tr}\left(\Delta\mathcal{M}_{12}\mathcal{M}_{22}^{-1}\mathcal{M}_{21}\right)
=T_{\text{E}}\frac{2r}{H^{2}}
\int_{0}^{\infty}\mathrm{d}\tilde{q}\int_{-\infty}^{\infty}\mathrm{d}\tilde{x}_{1}\,
\tilde{q}\Delta\tilde{\mathcal{M}}_{12}(\tilde{q};\tilde{x}_{1})
\tilde{\mathcal{M}}_{21}(\tilde{q};\tilde{x}_{1})
+\mathcal{O}\left(\frac{r^{2}}{H^{3}}\right) .\label{trace_eq_int_q_and_x}
\end{equation}
After the execution of the $\tilde{S}$ and $\tilde{T}$
integrals, the propagators are given by modified Bessel functions of the second kind,
\begin{eqnarray}
\Delta\tilde{\mathcal{M}}_{12}(\tilde{q};\tilde{x}_{1}) & = & 
\frac{1}{\pi}\frac{\tilde{q}}{\sqrt{1+\tilde{x}_{1}^{2}}}
K_{1}\left(\tilde{q}\sqrt{1+\tilde{x}_{1}^{2}}\right),\nonumber \\
\tilde{\mathcal{M}}_{21}(\tilde{q};\tilde{x}_{1}) & = & 
\frac{1}{2\pi}K_{0}\left(\tilde{q}\sqrt{1+\tilde{x}_{1}^{2}}\right).
\end{eqnarray}
This representation is suitable for a generalization to a uniaxially corrugated surface, as is done in the next section.

\section{Scalar Casimir-Polder potential for uniaxially corrugated surfaces}
\label{sec:corrugated}

We now extend the above method to uniaxially arbitrarily corrugated surfaces
$S_{1}$. For simplicity, we consider deformations along $x_{1}$ which can be
parametrized by a height function $h(x_{1})$ (overhangs could also be included
by an appropriate parametrization). The four-vector pointing onto the
structured surface reads $x=(x_{0},x_{1},x_{2},h(x_{1}))$.  The center of the
sphere is again located at $x_{1}=x_{2}=0$, $x_{3}=\bar{H}$. Here we write $\bar{H}$ instead of just $H$ to point out that it denotes the position of the sphere at a mean distance $\bar{H}$ from the surface. For corrugated surfaces the actual distance $H$ between the surface and the sphere is a function of the direction of corrugation $H=H(x_{1})$ (cf. also Fig. \ref{cpsine}). 
As for the planar situation considered earlier, it holds that $\bar{H}=H$. 

As the inverse correlator on the sphere does not change and the correlators
between the surfaces are still easy to compute, the main challenge for
extending the previous calculations to corrugated surfaces is the
determination of the inverse propagator $\mathcal{M}_{11}^{-1}$ on the
structured surface $S_{1}$. For the planar case, translational invariance
along the $\underline{x}$ directions allows for diagonalization in momentum
space and thus for an explicit solution as given in
Eq.(\ref{m11inverse_flach}).  By contrast, a structure on the surface breaks
translational invariance in the direction of corrugation and the functional
inverse of $\mathcal{M}_{11}$ is not known analytically in the general case.

Let us first derive a suitable representation of the problem. The desired
quantity $\mathcal{M}_{11}^{-1}$ is defined by the equation
\begin{equation}
\int_{\vec{x}\in S_{1}}\left.\mathcal{M}_{11}(\zeta;\vec{x}';\vec{x})
\mathcal{M}_{11}^{-1}(\zeta;\vec{x};\vec{x}'')=\delta(\vec{x}'-\vec{x}'')
\right|_{\vec{x}',\vec{x}''\in S_{1}} .\label{m11inv_bestimmungsgleichung}
\end{equation}
where $\mathcal{M}_{11}(\zeta;\vec{x}';\vec{x})$ is given by the Fourier
transform of \Eqref{greens_fct_for_del_square} with respect to the time
coordinate.  The integration over the surface in this case is defined
by
\begin{equation} 
\int_{\vec{x}\in  S_{1}}=\left.\int_{x_{1}}\int_{x_{2}}\sqrt{g(x_{1})}
\right|_{x_{3}=h(x_{1})},
\end{equation} 
with the determinant of the induced metric given by
$g(x_{1})=1+(\partial_{1}h(x_{1}))^{2}$.  Thus
\Eqref{m11inv_bestimmungsgleichung} explicitly reads 
\begin{equation}
 \int_{\vec{x}_{\parallel}}\sqrt{g(x_{1})}
\mathcal{M}_{11}(\zeta;\vec{x}_{\parallel}';\vec{x}_{\parallel})
\mathcal{M}_{11}^{-1}(\zeta;\vec{x}_{\parallel};\vec{x}_{\parallel}'')
=\frac{1}{\sqrt{g(x_{1}')}}\delta(\vec{x}_{\parallel}'-\vec{x}_{\parallel}''),
\label{m11inv_bg_mit_g}
\end{equation}
where $\vec{x}_\|$ summarizes the flat coordinates $x_1,x_2$ along the
two-dimensional surface.

In principle, we could try to solve this equation numerically and then plug in
the solution into the our formulas for the Casimir energy. However, due to the
singularity structure of this equation, and also since we rather need the
operator product $\Delta \mathcal{M}_{12}=\mathcal{M}_{11}^{-1}
\mathcal{M}_{12}$ than $\mathcal{M}_{11}^{-1}$ alone, we now multiply this
equation with $
\sqrt{g(x_{1}'')}\mathcal{M}_{12}(\zeta;\vec{x}_{\parallel}'';
\vec{x}_{\parallel}''',x_{3}''')$
from the left and integrate both sides of Eq. (\ref{m11inv_bg_mit_g}) over the
lateral coordinates $\vec{x}_{\parallel}''$ . One finds that
\begin{equation}
  \int_{\vec{x}_{\parallel}}\sqrt{g(x_{1})}
\mathcal{M}_{11}(\zeta;\vec{x}_{\parallel}';\vec{x}_{\parallel})
\Delta\mathcal{M}_{12}(\zeta;\vec{x}_{\parallel};
\vec{x}_{\parallel}''',x_{3}''')
=\mathcal{M}_{12}(\zeta;\vec{x}_{\parallel}';\vec{x}_{\parallel}''',x_{3}''')
  ,\label{bestimmunsgl_deltaM12_ortsr}
\end{equation} 
where $\Delta\mathcal{M}_{12}$ explicitly includes the metric factor of the
respective structure.  Since the above matrix elements are still diagonal in
$p_{2}$ in our uniaxial setup, we take the Fourier transform of
\Eqref{bestimmunsgl_deltaM12_ortsr} with respect to the 2-component. Slightly
renaming the coordinates, we finally get
\begin{equation}
\int_{x_{1}}\sqrt{g(x_{1})}\mathcal{M}_{11}(\zeta,p_{2};x_{1}';x_{1})
\Delta\mathcal{M}_{12}(\zeta,p_{2};x_{1};x_{1}'',x_{3}'')=
\mathcal{M}_{12}(\zeta,p_{2};x_{1}';x_{1}'',x_{3}'') .
\label{bestimmungsgl_deltaM12_unrescaled}
\end{equation}
This equation is indeed more suitable for a numerical integration. As a final
step, we again go over to dimensionless variables by rescaling all
dimensionful quantities by the distance parameter $H$.\footnote{Depending on 
the
relevant geometric parameters, a rescaling with a different parameter such as
a height amplitude or the wavelength of the corrugation may also be
useful.} Also introducing the combined momentum $q=\sqrt{\zeta^2+p_2^2}$
results in
\begin{equation}
\int_{\tilde x}\sqrt{g(\tilde x)}\tilde{\mathcal{M}}_{11}(\tilde
q;\tilde{x}';\tilde x)
\Delta\tilde{\mathcal{M}}_{12}(\tilde q;\tilde x)
=\tilde{\mathcal{M}}_{12}(\tilde q;\tilde x') ,
\label{bestimmungsgl_deltaM12_final}
\end{equation}
where we have dropped the coordinate subscript ``1'', since $x\equiv x_1$ is
the only relevant direction in this Green's function problem. The
dimensionless propagators $\tilde{\mathcal{M}}_{12}\equiv
\tilde{\mathcal{M}}_{21}$ and $\tilde{\mathcal{M}}_{11}$ are given by
\begin{eqnarray}
\tilde{\mathcal{M}}_{11}(\tilde q;\tilde{x}';\tilde x) & = &
\frac{1}{2\pi}K_{0}
\left(\tilde q\sqrt{(\tilde x'-\tilde x)^{2}+\left(\tilde{h}(\tilde{x}')-\tilde{
      h}(\tilde x )\right)^{2}}\right) , \label{m11_resc_general_height}\\
\tilde{\mathcal{M}}_{12}(\tilde q;\tilde x') & = & \frac{1}{2\pi}K_{0}
\left(\tilde q\sqrt{(\tilde x')^{2}+\left(\tilde h(\tilde{
      x}')-1\right)^{2}}\right) ,
\label{m12_resc_general_height}
\end{eqnarray}
where the metric factor reads
\begin{equation}
\sqrt{g(\tilde x)}=\sqrt{1+\left(\partial_{\tilde x} \tilde h(\tilde 
x )\right)^{2}},
\quad \tilde h(\tilde x) = \frac{1}{H}\, h( \tilde{x} H).
\label{metric_resc_general_height}
\end{equation}
Once \Eqref{bestimmungsgl_deltaM12_final} is solved for
$\Delta\tilde{\mathcal{M}}_{12}$, the solution can be plugged into the Casimir
energy formula. The scalar Casimir energy
between a sphere and a surface which is uniaxially corrugated along $x$ thus
yields in the limit $r\ll H$
\begin{equation}
E=-\frac{\hbar c }{2}\frac{r}{H^2}\, \alpha
+\mathcal{O}\left(\frac{r^{2}}{H^{3}}\right) ,\label{CP_energy_final}
\end{equation}
where
\begin{equation}
\alpha:= {2\int_{0}^{\infty}\mathrm{d}\tilde{
  q}\int_{-\infty}^{\infty}\mathrm{d}\tilde x\,\sqrt{g(\tilde x)}\tilde q
\Delta\tilde{\mathcal{M}}_{12}(\tilde q;\tilde
x)\tilde{\mathcal{M}}_{21}(\tilde q;\tilde x)}, \label{eq:alpha}
\end{equation}
is a dimensionless numerical factor that depends on the geometry parameters of
the configuration (measured in units of $H$). The Casimir-Polder limit is
obtained in the limit of the sphere radius $r$ being much smaller than any
other scale, $r\ll H,A,\lambda,\dots$, where $A$ denotes a typical amplitude
of the corrugation and $\lambda$ a typical corrugation wavelength. The factor
$\alpha$ thus is a function of $\alpha=\alpha(A/H, \lambda/H,\dots)$, but it is
independent of $r$.

From a technical perspective, the result of Eqs.~\eqref{CP_energy_final},
\eqref{eq:alpha} is very simple. It should be stressed that already the first
trace term in the initial Casimir-energy formula \eqref{E_tr_cycle} includes 
nine integrations for the trace: one over the imaginary frequency
and four times two integrations over the lateral surface coordinates. Due to
the trivial dependency of the integrand on the lateral coordinates of the
sphere in the Casimir-Polder limit, the number of integrations was then
reduced by four; moreover, the $n$ sum is just replaced by its first term in
this limit. The emerging translational invariance vertical to the direction of
corrugation reduces the number of integrations by another two. Thus -- instead
of nine -- we are left with three integrations: two of them are directly
visible in \Eqref{CP_energy_final}, the third one is needed for the
construction of $\Delta\mathcal{M}_{12}$ as a solution of
\Eqref{bestimmungsgl_deltaM12_final}.  These simplifications make the
Casimir-Polder limit accessible to numerical integration for arbitrary height
profiles.

The resulting two integrals in \Eqref{eq:alpha} are both convergent,
non-oscillatory and generically exhibiting a simple one-peak structure. 
The treatment of the singularity structure in the Green's functions equation
\eqref{bestimmungsgl_deltaM12_final}, however, requires some care and is treated in the Appendix.

\section{Sinusoidal corrugation\label{sec:sinus}}

\begin{figure}[h]
\begin{centering}
\includegraphics[scale=0.6]{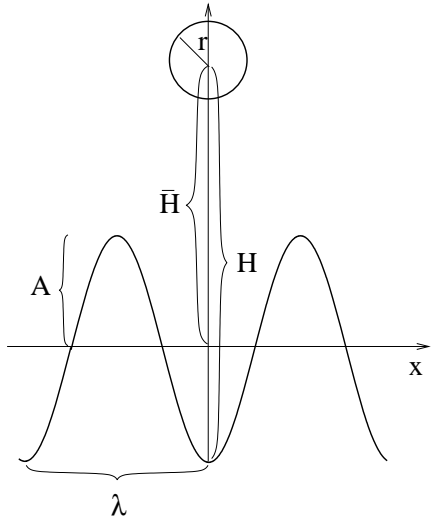}
\par\end{centering}

\caption{Sphere of radius $r$ at a mean distance $\bar H$ above a sinusoidally
  corrugated surface with amplitude $A$ and wavelength
  $\omega=\frac{2\pi}{\lambda}$. In our conventions, we fix the sphere at the
  lateral coordinate $x=0$, and effectively shift the structure function
  $h(x)$ by varying the phase $\phi$. In this plot, we have set $\phi=-\pi/2$. The
  distance parameter $H$ measures the sphere-surface distance along the global
  normal, such that $H=0$ corresponds to sphere-surface contact for all values
  of $\phi$. }

\label{cpsine}
\end{figure}

As a first nontrivial example, let us calculate the scalar Casimir-Polder
potential for a sinusoidal corrugation, see Fig. \ref{cpsine}. The potential
for this structure is given by Eqs.~\eqref{CP_energy_final}, \eqref{eq:alpha},
where we use $h(x)=A\sin(\omega x+\phi)$ as a height function appearing in the
propagators \Eqref{m12_resc_general_height} as well as in the surface metric
factor of \Eqref{metric_resc_general_height}. The phase $\phi$ is used to
modulate the relative position of the structure beneath the sphere, since the latter is
always fixed at $x=0$. The mean distance between surface and atom is denoted
by $\bar H$, whereas $H$ characterizes the distance of the atom to the surface
along the global surface normal. Hence $H$ can be viewed as a function of
$\phi$ in our conventions, $H=H(\phi)$ with $H(-\pi/2)=\bar H +A$
at the sine minimum and $H(\pi/2)=\bar H -A$ at the sine maximum, i.e., $H=0$
always corresponds to atom-wall contact, where the limit $r\ll H$ is implicitly understood.

As the crucial building block for the Casimir-Polder potential, we solve the
Green's function equation \Eqref{bestimmungsgl_deltaM12_final} numerically on
a 1-D lattice in $x$ direction. This requires to invert the propagator
$\tilde{\mathcal{M}}_{11}$ on the corrugated surface. Even though the
singularity of this propagator at coincident points is integrable in the
continuum, the discretized version needs to deal with this singularity
explicitly. This is done by introducing a regularization parametrized by a
short-distance cutoff $\epsilon$, which can be removed after the continuum
limit has been taken. Details of how this procedure is implemented numerically
are given in Appendix \ref{sec:details}.

In the following, we display our results for the Casimir-Polder energy always
normalized with respect to the planar-surface case (for consistency, the
normalization factor is also determined numerically) . In this manner, the
geometry-induced effects are better visible. Furthermore, we expect that these results
for the scalar case give a qualitative estimate also for the electromagnetic
case for which the normalizing prefactor has a different distance dependence.

\begin{figure}[h]
\begin{centering}
\includegraphics[scale=0.9]{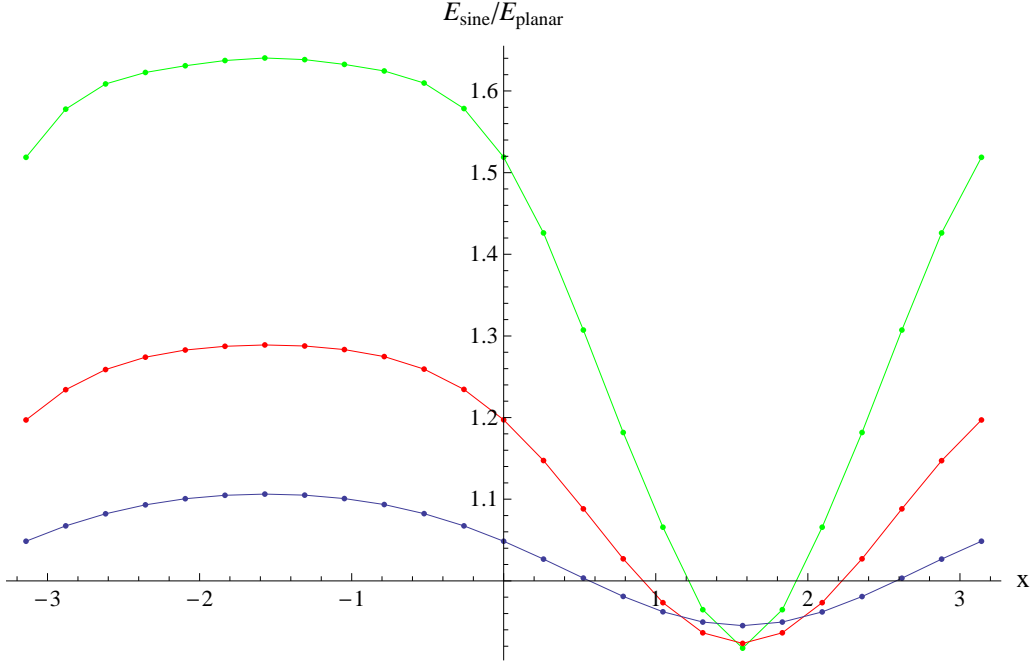}
\par\end{centering}

\caption{We give the ratio $E_{\text{sine}}/E_{\text{planar}}$ for three
  different separations $\bar{H}/A=4$ (blue) $\bar{H}/A=2$ (red) and $\bar H/A=1.25$
  (green), from bottom to top. The frequency of the sine structure is kept
  fixed at $\omega A=1$. Above the corrugation trough at
  $\phi=-\frac{\pi}{2}$, it holds that $E_{\text{sine}}/E_{\text{planar}}>1$,
  since the influence of the concavities of the corrugation towards the sphere
  is not accounted for by $E_{\text{planar}}$. Similarly, one finds
  $E_{\text{sine}}/E_{\text{planar}}<1$ above the maximum of the corrugation
  at $\phi=\frac{\pi}{2}$. For larger ratios of $\bar H/A$, i.e. larger
  distances, the result approaches the planar case. Note that even at larger
  separations $\bar H/A=4$, a pure planar approximation deviates from its true
  value by up to $10\%$.}

\label{horizontal_deviation}
\end{figure}

In Fig. \ref{horizontal_deviation}, we plot
$E_{\text{sine}}/E_{\text{planar}}$ as a function of the horizontal position
of the sphere above the sinusoidal corrugation between $\phi=-\pi$ and
$\phi=\pi$ for three different mean separations $\bar H/A=4,2,1.25$. The
corrugation frequency is chosen to be $\omega A$=1, all units are set by the
corrugation amplitude $A$. As expected, above the corrugation minimum, e.g.,
at $x=-\pi/2$, the Casimir-Polder potential lies above the planar estimate
since the plate bends towards the sphere. Analogously, above the corrugation
maximum at $\pi/2$, $E_{\text{sine}}/E_{\text{planar}}<1$. It is clear that
the ratio $E_{\text{sine}}/E_{\text{planar}}\rightarrow1$ as $\bar
H/A\rightarrow\infty$, since the corrugation cannot be resolved anymore for
greater distances. However, it is quite noticeable that the deviation from the
planar case is still up to $10\%$ even at large separations $\bar H/A=4$.

\begin{figure}[h]
\begin{centering}
\includegraphics[scale=0.8]{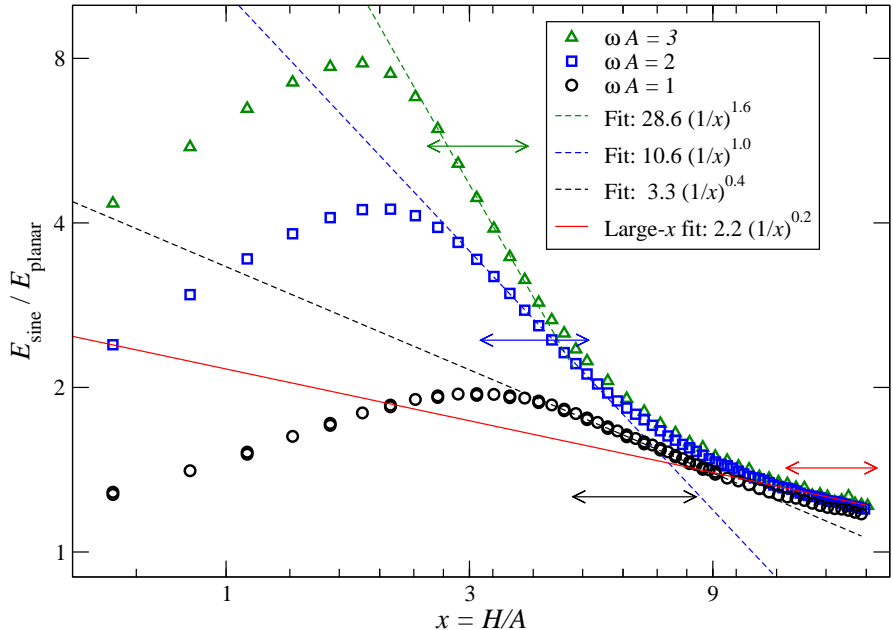}
\par\end{centering}

\caption{Normalized Casimir-Polder energy $E_{\text{sine}}/E_{\text{planar}}$
  above a corrugation minimum $\phi=-\pi/2$ versus the normalized distance
  $x\equiv H/A$ for three different corrugation frequencies $\omega A=1,2,3$,
  respectively. All units are set by the corrugation amplitude $A$. Small
  distances are governed by a linear increase with anomalous dimension
  $\eta=-1$, cf. \Eqref{eq:eta}. The drop-off beyond the peak is characterized
  by an $\omega$-dependent anomalous dimension $\eta\simeq 0.4,1.0,1.6$ for
  $\omega A=1,2,3$. At larger distances $H/A\sim 10$, all normalized energies
  approach a universal curve with $\eta\simeq 0.2$. The corresponding fit
  regions are indicated by horizontal arrows. Also the increase towards the
  peak can be parametrized by a power-law with anomalous dimensions
  $\eta\simeq- 0.33, -0.57, -0.67$ for $\omega A=1,2,3$ (not shown in the plot).}

\label{HfitSineWell}
\end{figure}

In Fig. \ref{HfitSineWell}, we display $E_{\text{sine}}/E_{\text{planar}}$ as
a function of the vertical position of the sphere above a minimum of the
corrugation ($\phi=-\pi/2$) for different corrugation frequencies $\omega
A=1,2,3$. In the limits $H/A\rightarrow\infty$ and $H/A\rightarrow0$, we find
that $E_{\text{sine}}/E_{\text{planar}}\rightarrow1$. This is expected, since
in the first limit the corrugation of the plate cannot be resolved as it is
too small compared to the distance. In the second limit, the corrugation is
irrelevantly large compared to the distance, i.e. the sphere does not notice it locally. In the region where $H\sim A$,
the potential for the corrugated surface clearly deviates from the
corresponding planar case. One can see that the effect becomes more pronounced
for larger corrugation frequencies, i.e. shorter surface periodicity.

We identify various regimes which can be classified in terms of an anomalous
dimension $\eta$ which measures the deviation of the Casimir-Polder potential
from the planar case, 
\begin{equation}
E_{\text{corrugation}}\sim \frac{1}{H^{2+\eta}}, \label{eq:eta}
\end{equation}
with $\eta=0$ for the planar case. At small distances, $H/A\ll1$, we find a
linear increase of the normalized potential
$E_{\text{sine}}/E_{\text{planar}}$ with $H/A$,  implying an anomalous
dimension of $\eta=-1$. A linear fit to the short-distance data (not shown in
Fig.~\ref{HfitSineWell}) in the well yields
$E_{\text{sine}}/E_{\text{planar}}\simeq 1+\beta (H/A)$. The linear
coefficient $\beta$ depends on the frequency, $\beta=\beta(\omega/A)\simeq
0.5, 2.3, 5.2$ for $\omega A=1,2,3$; within the numerical accuracy, this
dependence is compatible with a power law $\beta\sim (\omega A)^2$. 

At larger distances $H/A\sim\mathcal{O}(1)$, the normalized energy develops a
peak. Various regimes can be identified near the peak and also in the drop-off
region. The increase towards the peak as well as the decrease right beyond the
peak can be characterized by power laws parametrized by an $\omega$-dependent
anomalous dimension. Towards the peak, we find $\eta\simeq -0.33, -0.57, -0.67$
for $\omega A=1,2,3$, and the fit beyond the peak yields $\eta\simeq 0.4, 1.0,
1.6$ for $\omega A=1,2,3$. For even larger distances near $H/A\simeq 10$, we
observe that all normalized energies approach a universal curve being
characterized by an anomalous dimension $\eta=0.2$; in particular, the
anomalous dimension shows no sizeable $\omega$ dependence anymore.

Whereas this observation might come as a surprise in the present formalism, it
can easily be interpreted in the framework of the worldline picture of the
quantum vacuum \cite{Gies:2003cv}. In this picture, quantum fluctuations are
mapped onto random paths characterizing the spacetime trajectories of these
fluctuations. In order to contribute to the Casimir interaction energy, such a
trajectory has to intersect with both surfaces, the sphere and the corrugated
plate in the present case. This implies that the fluctuation has an average
extent of the order of the surface separation $H$. Due to isotropy of the
vacuum fluctuations, the relevant worldlines also have a lateral extent of
this order. This implies that the fluctuation integral also averages over
structures of the corrugation which are smaller than $H$. Higher corrugation
frequencies with $\omega H\gg 1$ thereby become irrelevant for the
Casimir-Polder potential, as is demonstrated by the universal drop-off for
larger $H/A$. 

For even larger distances $H/A\gg 10$, the power law cannot continue
for arbitrarily large $H/A$, since the Casimir-Polder potential eventually has to approach
the planar limit. In this large-distance regime, we have only a few reliable
numerical data points, as the discretization artifacts increase, once the
lattice spacing approaches the corrugation wavelength. The available data is
compatible with a logarithmic approach towards
$E_{\text{sine}}/E_{\text{planar}}\to 1$ for $H/A\to \infty$.

\begin{figure}[h]
\begin{centering}
\includegraphics[scale=0.8]{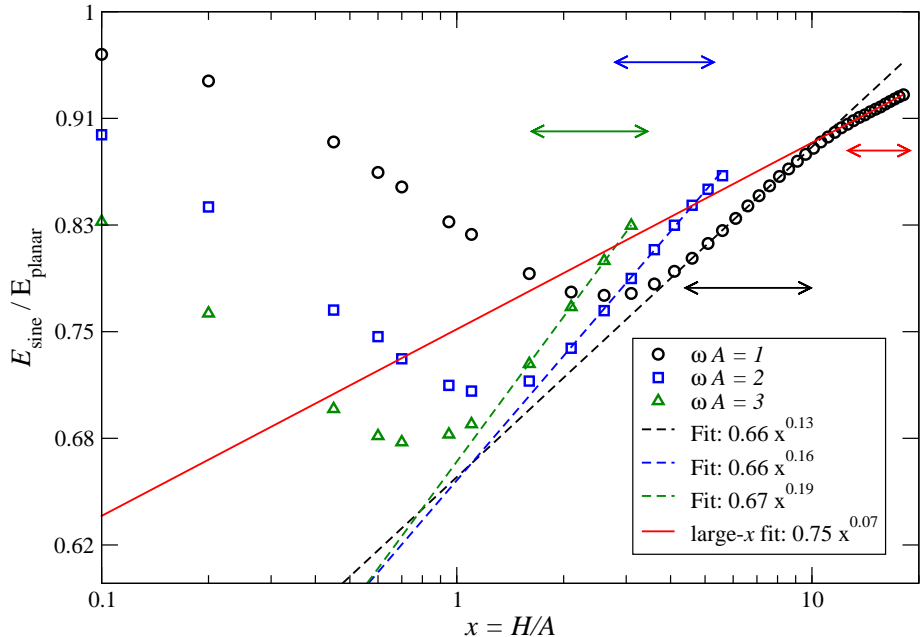} 
\par\end{centering}

\caption{Normalized Casimir-Polder energy $E_{\text{sine}}/E_{\text{planar}}$
  above a corrugation maximum $\phi=\pi/2$ versus the normalized distance
  $x\equiv H/A$ for three different corrugation frequencies $\omega A=1,2,3$,
  respectively. All units are set by the corrugation amplitude $A$. The
  increase beyond the dip is characterized by an $\omega$-dependent anomalous
  dimension $\eta\simeq -0.13,-0.16,-0.19$ for $\omega A =1,2,3$. At larger
  distances $H/A\gtrsim 10$, a power law  with $\eta\simeq -0.07$ is observed
  for the $\omega A=1$ curve. The corresponding fit regions are indicated by
  horizontal arrows. Also the decrease towards the
  dip can be parametrized by a power-law with anomalous dimensions
  $\eta\simeq 0.09, 0.11, 0.11$ for $\omega A=1,2,3$ (not shown in the plot).}

\label{HfitSineHill}
\end{figure}

Finally, we compute the Casimir-Polder potential above a maximum of the sine
structure at $\phi=+\pi/2$. As expected, the Casimir-Polder energy is always
smaller than in the planar case as the surfaces bends away from the atom and
approaches the planar result in the two limits $H/A\to 0$ and $H/A\to \infty$,
see Fig.~\ref{HfitSineHill}. Starting from an initial decrease of the
normalized energy for small distances $H/A$, a power-law decrease develops
towards the dip with $\eta\simeq 0.09,0.11,0.11$ for $\omega A =1,2,3$. Beyond
the dip near $H/A\sim 1$, a power-law increase follows with anomalous
dimension $\eta =-0.13,-0.16,-0.19$ for $\omega A=1,2,3$, respectively. Again,
we observe a linear $\omega$ dependence of $\eta$ in this regime. Also, a
second power-law regime is found for larger distances $H/A\gtrsim 10$ with an
anomalous dimension $\eta =-0.07$ for the $\omega A=1$ data. Due to an
increase of the discretization artifacts, no reliable data for larger $\omega$
is available, such that the expected universality in this distance regime
still needs to be shown.

\section{Sawtooth corrugation \label{sec:sawtooth}}

As a second example, we study the Casimir-Polder potential for a sawtooth
structure, where the wavelength $\lambda$ is 2.8 in terms of the amplitude
$A$, i.e. the dominant frequency of its Fourier decomposition is $\omega A
\simeq 0.45$. These parameters reflect the specifications of a sawtooth
structure used in a recent experiment \cite{Warring}.
%Reference to experiment?
For practical purposes, we actually use a smoothed, continuous sawtooth-like
structure function with wavelength $\lambda$, starting at $h(0)=0$, rising
linearly to its maximum amplitude $A$ at $h(0.8\lambda)$ and dropping linearly
to zero again at $h(\lambda)=h(0)$.

\begin{figure}[h]
\begin{centering}
\includegraphics[scale=0.8]{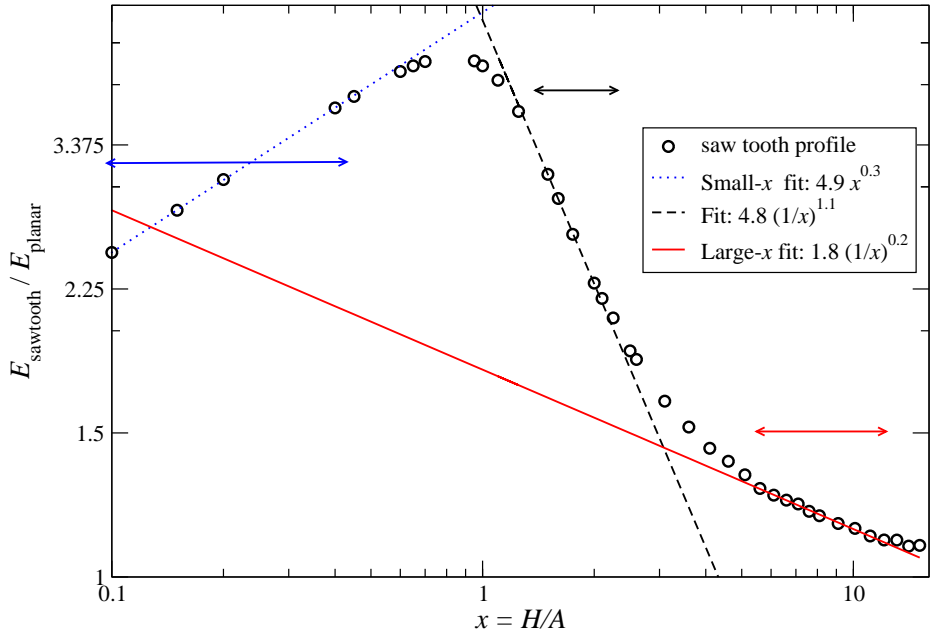} 
\par\end{centering}

\caption{Normalized Casimir-Polder energy
  $E_{\text{sawtooth}}/E_{\text{planar}}$ above a corrugation minimum of the
  saw-tooth profile (see text) versus the normalized distance $x\equiv H/A$
  for $\omega A\simeq 0.45$. All units are set by the corrugation amplitude
  $A$. The increase towards the peak is governed by a power-law with anomalous
  dimension $\eta=-0.3$. The drop-off beyond the peak is characterized by an
  $\omega$-dependent anomalous dimension $\eta\simeq 1.1$. At larger distances
  $H/A\sim 10$, the normalized energy approaches a curve similar to the curves
  for the sine structure with $\eta\simeq 0.2$. The corresponding fit regions
  are indicated by horizontal arrows. }

\label{sawtooth}
\end{figure}

In Fig. \ref{sawtooth}, we plot $E_{\text{sawtooth}}/E_{\text{planar}}$ above the
corrugation minimum. Qualitatively, the result is similar to the sine
structure and reveals the various analogous regimes. Quantitatively, the peak
and consequently some of the anomalous dimensions are more pronounced. The
increase towards the peak follows a power-law with anomalous dimension
$\eta\simeq-0.3$. For the decrease right beyond the peak at $H/A\gtrsim 1$, we
find an anomalous dimension of $\eta\simeq 1.1$. At larger distances $H/A\sim
10$, we again observe a second power law with anomalous dimension $\eta\simeq
0.2$ which agrees quantitatively with the anomalous dimension in the
sinusoidal case.

Within the worldline picture of quantum fluctuations discussed above, this
agreement can immediately be understood from the fact that the fluctuation
integrals again average over the corrugation structures small compared to the
distance parameter $H$.

\section{Conclusions\label{sec:conclusions}}

In this work, we have presented a new approach to Casimir-Polder forces for
corrugated surfaces which does not rely on a perturbative ordering of length
scales. Our approach is based on the constrained-functional-integral approach
\cite{Bordag:1983zk} which -- for uniaxial corrugations -- boils down to a
one-dimensional Green's function equation along the corrugation
direction. This equation is accessible to direct numerical integration
provided the integrable singularity structures are carefully taken into
account. For simplicity, we have studied the Dirichlet scalar analogue of the
electromagnetic Casimir-Polder case, defined by a fluctuating scalar field
satisfying Dirichlet boundary conditions on the surface and the ``atom''; the
latter is modeled by a small sphere in the limit of zero radius in our
approach.

Two periodic corrugations, a sine and a sawtooth function, are studied in
detail, revealing various regimes with distinct distance dependencies. For a
classification of these distance dependencies, we have introduced the notion
of an anomalous dimension characterizing the deviation of the distance power
law of the Casimir-Polder potential from the planar case.  In particular, the
larger-distance behavior $H/A\gtrsim 1$ exhibits two different power-law
behaviors with positive anomalous dimensions above a corrugation minimum and
with negative anomalous dimensions above a corrugation maximum. In either
case, the intermediate distance behavior near $H/A\sim 1$ is characterized by
an anomalous dimension, the modulus of which increases linearly with the
corrugation frequency.

Most importantly, we have identified a larger-distance regime near $H/A\sim
10$ where all data above a corrugation minimum is characterized by a universal
anomalous dimension $\eta\simeq 0.2$. This anomalous dimension still depends
on the position above the corrugation, e.g., $\eta\simeq0.07$ above a minimum,
but we have found no dependence neither on the shape of the periodic profile
nor on the frequency as long as $\omega H\gg 1$. Within the worldline picture
of the quantum vacuum where fluctuation averages are mapped onto random-path
averages, this universality can be understood from the fact that small-scale
structures are averaged out and become irrelevant at larger distances. This
observation also justifies to use the notion of universality and anomalous
dimensions, since the fluctuation averages are reminiscent to those of
critical phenomena. The resulting Casimir-Polder potential can be viewed as a
``renormalized'' effective Hamiltonian where the running IR cutoff is provided
by the atom-wall distance.

We would like to stress that this universality as well as the nontrivial
power-law behavior cannot be deduced from a perturbative analysis, since
perturbation theory in the height profile is a Taylor expansion in powers of
$A/H$, whereas a nontrivial anomalous dimension $\eta\in \mathbbm{R}$
corresponds to a $(A/H)^\eta$ dependence of the Casimir-Polder law. Therefore,
the development and use of a nonperturbative method was absolutely crucial for
this work.

Even though we only considered the Dirichlet scalar case, the
constrained-functional-integral formalism can straightforwardly be extended to
the electromagnetic case as well \cite{Emig:2003eq} which carries over to a
straightforward generalization of our techniques. Since no monopole
fluctuations exist in the electromagnetic case due to charge conservation, the
leading-order Casimir-Polder potential in the planar case follows a
$\sim1/H^4$ distance law instead of $\sim1/H^2$ in the present case. But apart
from this, we do not expect further dramatic differences as far as the
corrugation-dependencies are concerned. Therefore, the Dirichlet scalar case
may be taken as a rough qualitative estimate also for the electromagnetic
case; in particular, we expect the occurrence of anomalous dimensions of the
same order of magnitude. 

The notion of anomalous dimension is also of direct use for Casimir-Polder
experiments based on quantum reflection such as the atom-beam spin echo
technique \cite{DeKieviet1, DeKieviet2, DeKieviet3}. In a certain sense, such an experiment measures the
local shape of a potential and thus is directly sensitive to anomalous
dimensions. Indeed, the results of recent measurements with corrugated
surfaces can be parametrized by anomalous dimensions of order one. Of course,
a direct comparison between our results and those of an experiment requires
much more than the computation of anomalous dimensions, since the atoms near
the wall can move into all directions and not only along the global normal. The
full Casimir-Polder potential needs to be mapped out, and the time-dependent
quantum reflection problem in this potential has to be solved. In any case,
the approach presented here lays the foundation for this future program. 

\section*{Acknowledgments}
The authors gratefully acknowledge useful discussions with T.~Emig and
R.L.~Jaffe. HG thanks the DFG for support under grant No. GI~328/5-1
(Heisenberg program). 
\appendix

\section{Numerical procedure\label{sec:details}}

In the following, we detail our implementation for the numerical evaluation of
the Casimir-Polder potential for arbitrary uniaxial corrugations, cf. sections
\ref{sec:sinus} and \ref{sec:sawtooth}.  We proceed as follows: First, we
solve the Green's function equation for the associated propagator
$\Delta\mathcal{M}_{12}$ Eq.\eqref{bestimmungsgl_deltaM12_final} by
discretizing the equation with respect to the spatially lateral coordinate
$x$.  The result is then plugged into \Eqref{CP_energy_final}, yielding the
Casimir-Polder energy upon integration of $\tilde x$ and $\tilde q$.

For the first step, we introduce two parameters: $\pm L_{x}$ which labels the
left and right cutoff of the spatial integration, and $N_{x}$ denoting the
number of spatial discretization sites, respectively.  In the end, we remove
the discretization by a continuum extrapolation $N_{x}\rightarrow \infty$.

In principle, $L_x$ is a physical parameter encoding the physical size of the
surface. Here, we will not make use of this option of studying finite-size
effects, but compute the Casimir-Polder potential in the ideal infinite
surface limit by extrapolating to $L_x\to\infty$. For this, we fix the
position of the sphere above the plate at $x=0$ and choose a symmetric cutoff
for $x\in[-L_x,L_x]$. The two limits, continuum ($N_x\to \infty$) and
infinite-length ($L_x\to\infty$) limit, have to be taken such that the lattice
spacing $a_x=2L_x/N_x$ also goes to zero, $a_x\to0$. This can be ensured by
choosing a suitable function $L_x=L_x(N_x)$, satisfying
$L_x(N_x\to\infty)\to\infty$ and $L_x(N_x)/N_x\to 0$ as $N_x \to \infty$. In practice, we use 
\begin{equation}
L_x(N_x)=\frac{a_{0x}}{2} \, \sqrt{N_x N_{0x}},\label{eq:L_x}
\end{equation}
where $a_{0x}$ defines a reference lattice spacing at a reference site number
$N_x=N_{0x}$. Note that the lattice spacing $a_x\equiv a_x(N_x)= 2L_x(N_x)/N_x
= a_{0x} \sqrt{N_{0x}/N_x}$ goes to zero in the continuum limit $N_x\to
\infty$, while $L_x\to \infty$ approaches the infinite length limit. Therefore
all these idealized limits are controlled by one parameter: $N_x$. In
practice, the finite-length corrections have always been found to be small
compared to discretization effects. In general, it suffices to choose the
reference lattice spacing such that typically $L_x(N_{0x})=2 H$, where
$N_{0x}$ specifies the coarsest lattice in the calculation.

One serious complication arises when discretizing
Eq.\eqref{bestimmungsgl_deltaM12_final}: due to the pole of the zeroth Bessel
function $K_0$ at its origin, the matrix $\mathcal{M}_{11}^{ij}$ that emerges upon
the discretization of the spatial arguments diverges in its diagonal entries,
i.e., for the case when the spatial discretization sites lie on top of each
other. Whereas these divergencies are integrable when solving the problem in
the continuum, the discretized matrix becomes singular. Therefore, a
regularization procedure is required that facilitates to first take the
continuum limit before the regulator can safely be removed. Here, we use a UV
regularization for the propagator in Eq. \eqref{m11_resc_general_height} for small
arguments $z$ controlled by a small parameter $\epsilon$:
\begin{equation}
\mathcal{M}_{11}(z)=\begin{cases}
\frac{1}{2\pi}K_{0}(z) & ,z\leq\epsilon\\
-\frac{1}{2\pi}(\ln(z+\epsilon)-K_{0}(\epsilon)-\ln(2\epsilon)) &
,z>\epsilon\end{cases} ,\label{m11_reg_ln}
\end{equation} 
where $z$ summarizes all arguments of the propagator including both spatial
and momentum contributions, entering the Bessel function as a single argument,
cf. \Eqref{m11_resc_general_height}. The physical result is expected to arise
in the limits $N_{x}\rightarrow \infty$ and $\epsilon\rightarrow 0$ with the
continuum limit to be taken first before the regulator is removed.

In a numerical calculation where $N_x$ and $\epsilon$ are always finite, the
order of limits done by extrapolation requires a careful choice of $N_x$ and
$\epsilon$. It is already intuitively clear that smaller values of $\epsilon$
require larger values of $N_x$, since the proper resolution of a more
pronounced singularity for smaller $\epsilon$ needs a finer lattice.  As the
pole in the inverse propagator on the corrugated surface $S_{1}$ persists
irrespectively of the corrugation, the numerical discretization and
regularization errors can be tested in the planar situation where the
analytical result is known (cf. section \ref{sec:planar}): there, the
dimensionless factor $\alpha$ amounts to $\frac{1}{4\pi}$.

\begin{figure}[h]
\begin{centering}
\includegraphics[scale=0.9]{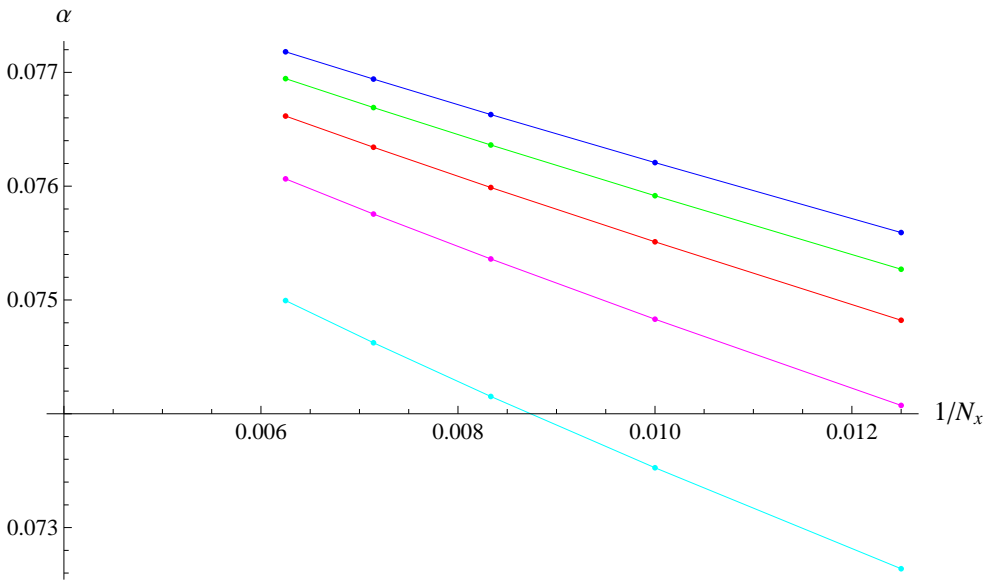}
\par\end{centering}

\caption{Numerical values for the dimensionless contribution of the
  lowest-order trace term $\alpha$ \eqref{eq:alpha} in the plane-sphere 
  configuration as a function of the inverse number of discretization sites
  $N_{x}$ for five values of the cutoff parameter $\epsilon$,
  $\epsilon=5\cdotp10^{-5},2\cdotp10^{-4},4\cdotp10^{-4},6\cdotp10^{-4},8\cdotp10^{-4}$
  from bottom to top. The analytical value for $\alpha$ is $1/(4\pi)\approx
  0.07958$. For fixed $\epsilon$, the result scales linearly with the
  discretization $1/N_{x}$ to a good approximation and appears to converge
  with $1/N_{x}\rightarrow 0$, but it is also visible that the gradients of the curves grow
  as $\epsilon\rightarrow0$.}

\label{traceimNraum}
\end{figure}

In Fig. \ref{traceimNraum}, we plot $\alpha$ as a function of the inverse
number of discretization sites $1/N_x$ for different values of the cutoff
$\epsilon$ in the planar case. The values for $\alpha$ depend linearly on
$1/N_{x}$ to a good approximation and appear to converge for different cutoffs
$\epsilon$ as $1/N_{x}\rightarrow 0$.

\begin{figure}[h]
\begin{centering}
\includegraphics[scale=0.9]{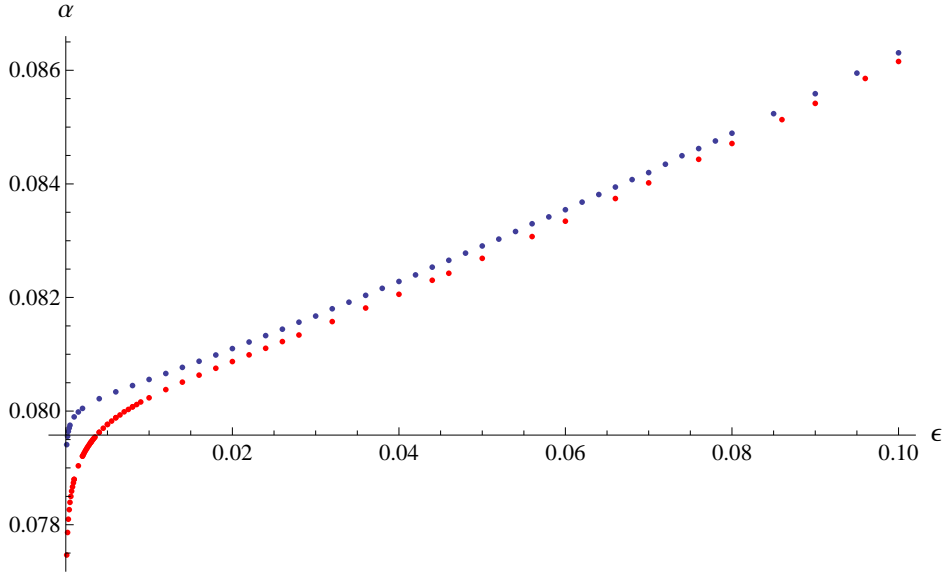}
\par\end{centering}

\caption{ Continuum limit for $\alpha$ as defined in Eq. \eqref{eq:alpha} after linear
  extrapolation to $1/N_{x}\rightarrow 0$ (using $1/N_{x}=80$ and
  $1/N_{x}=100$  in the lower (red) and $1/N_{x}=180$ and
  $1/N_{x}=200$ data in the upper (blue) curve) as a function of the cutoff parameter $\epsilon$ for
  values of $10^{-4}<\epsilon<0.1$. The intersection between the two plot
  axes is chosen at $\alpha=\frac{1}{4\pi}$, which is the exact value for
  $\alpha$ in the planar case. The two curves can both be separated into several regimes. Consider e.g. the lower curve: It holds that for very small values of $\epsilon$
  ($\epsilon\lessapprox0.0045$), the result for $\alpha$ lies below the
  analytical value and appears to diverge as $\epsilon\rightarrow0$, since the
  integrable singularity is not resolved by the number of sites $N_{x}$ used
  for the extrapolation. In the region
  $0.01\lessapprox\epsilon\lessapprox0.04$, $\alpha(\epsilon$) grows linearly
  with $\epsilon$ (for $\epsilon\gtrapprox0.04$, we identify a
  $\alpha(\epsilon)\sim\epsilon^2$ dependence). An extrapolation
  $\alpha(\epsilon\rightarrow0)$ in the range
  $0.01\lessapprox\epsilon\lessapprox0.04$ thus provides an estimate for the
  cutoff-independent value of $\alpha$ (cf. \Eqref{alpha_series}). For the upper, blue curve the respective regimes can also be identified. However, since the resolution of the structure is increased by a greater number of $N_{x}$, the important linear regime is shifted to lower $\epsilon$-values.}

\label{traceimepsraum}
\end{figure}

Next, we extrapolate the values for $\alpha$ linearly to $1/N_{x}=0$; as the
linearity persists to a good approximation for all values of $N_x$ in
Fig.~\ref{traceimNraum}, it suffices to use only two data points for the extrapolation. We give two separate extrapolations for $N_{x}=80$, $N_{x}=100$ and $N_{x}=180$
, $N_{x}=200$, respectively. The result is plotted as a function of
$\epsilon$ in Fig. \ref{traceimepsraum}. Recall that the analytical value for
$\alpha$ yields $1/(4\pi)\approx 0.07958$ for the flat plate, which is chosen
to be exactly the origin of the coordinate system in
Fig. \ref{traceimepsraum}.

The graphs in Fig. \ref{traceimepsraum} can in fact be divided into several
regions. Consider, e.g. the lower curve: For values of $\epsilon\lessapprox0.0045$, the extrapolation
$1/N_{x}\rightarrow 0$ underestimates $\alpha$ and even appears to diverge as
$\epsilon \rightarrow 0$. This agrees with our expectation that the integrable
singularity in the Green's function equation has not been properly resolved
with the underlying discretization; higher values of $N_x$ would be required
for a more reliable estimate. This small-$\epsilon$ branch therefore
corresponds to a region in parameter space where the result arising from the
correct order of limits (first $N_x\to \infty$, then $\epsilon\to0$) is not
yet visible. 

At about $\epsilon\gtrapprox0.01$, $\alpha$ exhibits a clear linear growth
with $\epsilon$. For $\epsilon\gtrapprox0.04$, higher power corrections become
visible.  We conclude that the
cutoff-dependent factor $\alpha(\epsilon)$ can well be approximated by a power
series above the value of $\epsilon\gtrapprox0.01$,
\begin{equation}
\alpha(\epsilon)=\alpha_{0}+\alpha_{1}\epsilon+\alpha_{2}\epsilon^2+\dots\
.\label{alpha_series}  
\end{equation}
Thus, by extrapolating the values for $\alpha$ to $\epsilon=0$ in the region
where $\alpha$ grows linearly with $\epsilon$, we obtain a cutoff-independent
result $\alpha_{0}$.  As for the extrapolation $1/N_{x}\rightarrow 0$, it
suffices to use only two sites in $\epsilon$ in the linear regime to extract
$\alpha_{0}$; of course, also more data points for a higher polynomial fit
could easily be employed at the expense of computing time.

From
Fig.~\ref{traceimepsraum}, we identify for the $1/N_{x}=80$ and
  $1/N_{x}=100$ data
$0.01\lessapprox\epsilon\lessapprox0.04$ as the region where
$\alpha(\epsilon)$ grows linearly with $\epsilon$ with only very small
higher-power corrections. Choosing the data points at $\epsilon=0.02$ and
$\epsilon=0.025$ for a linear extrapolation, we obtain $\alpha_{0}=0.07970$
which nicely matches the analytical value, the error being below
1\%.\footnote{For the study of corrugated surfaces, we have carefully studied
  whether the interval linear in $\epsilon$ is shifted and the extrapolation
  has to be adjusted accordingly. It turns out that the endpoints of the
  linear region are indeed slightly shifted for structured surfaces, but the
  sampling points $\epsilon=0.02$ and $\epsilon=0.025$ have always been in the
  linear region for all examples.}  As a check of the continuum limit, an
extrapolation using $N_{x}=180$ and $N_{x}=200$ sites (upper curve in Fig.~\ref{traceimepsraum},again at
$\epsilon=0.02$ and $\epsilon=0.025$) yields $\alpha_{0}=0.0799554$, which is
also within $1\%$ of the analytical value. The small deviations between these
two results can be taken as a measure for the overall numerical uncertainty.
One can see, that choosing larger values of $N_x$ for the continuum extrapolation
also results in an extension of the linear $\epsilon$ regime to smaller
$\epsilon$ values.

It should be mentioned that the choice of required $N_x$ values also depends
on the corrugation parameters. For instance for high values of the corrugation
frequency, a better resolution is needed; as a rule of thumb, the lattice
spacing $a_x$ should always be smaller than the smallest dominant wave length
of the corrugation. 

All numerical calculations for this work have been performed on a standard
desktop computer with standard linear algebra packages. Depending on the
discretization, the calculation of a typical data point including continuum
limit and regulator removal takes on the order of seconds to several minutes. 
Since the linear-algebra routines scale with $\sim N_x^3$, the computational
cost for very fine discretizations can rapidly increase.

\end{document}